\documentclass[10pt, conference]{IEEEtran}

\usepackage{amsmath, amssymb, enumerate}
\usepackage{fancybox}
\usepackage{amsfonts}
\usepackage{algorithm}
\usepackage{algorithmicx}
\usepackage{algpseudocode}
\usepackage[usenames]{color}
\usepackage{listings}
\usepackage{graphicx,dblfloatfix}
\usepackage{times}
\usepackage{psfrag}
\usepackage{subfigure}
\usepackage{caption}
\usepackage{multirow}
\usepackage{epsfig}
\usepackage{url}
\usepackage{color}
\def\fixme#1{\typeout{FIXED in page \thepage : {#1}}
\bgroup \color{red}{[FIXME: {#1}]} \egroup}
\usepackage{rotating,tabularx}

\begin{document}
\title{A Covert Channel Using Named Resources}

\author{\IEEEauthorblockN{Joshua Davis}
\IEEEauthorblockA{Email: namecc@covert.codes}

\and

\IEEEauthorblockN{Victor S. Frost}
\IEEEauthorblockA{University of Kansas\\ Email: frost@ku.edu}
}

\maketitle
\thispagestyle{empty}

\begin{abstract}
A network covert channel is created that uses resource names such as addresses to convey information, and that approximates typical user behavior in order to blend in with its environment.  The channel correlates available resource names with a user defined code-space, and transmits its covert message by selectively accessing resources associated with the message codes.  In this paper we focus on an implementation of the channel using the Hypertext Transfer Protocol (HTTP) with Uniform Resource Locators (URLs) as the message names, though the system can be used in conjunction with a variety of protocols.  The covert channel does not modify expected protocol structure as might be detected by simple inspection, and our HTTP implementation emulates transaction level web user behavior in order to avoid detection by statistical or behavioral analysis.
\end{abstract}

\section{Introduction}

Covert channels are essentially inter-process communication mechanisms designed to evade security systems that may otherwise monitor or prevent communication between the endpoint processes.  Network covert channels have existed for decades \cite{Girling87}, and have increased in complexity, diversity, and capability over time \cite{Ahmadzadeh13}, \cite{Luo12},  \cite{Mazurczyk13}, \cite{Gasior11}.  As these channels advanced, so did technology used to detect or disrupt them \cite{Gilbert09}, \cite{Tumoian05}, \cite{Cabuk04}, \cite{Cabuk09}, \cite{Gianvecchio11}, \cite{Giles02}, and whereas early network covert channels may have operated, for example, by simply placing data in unused header fields, the requisite cleverness necessary to remain undetected in security-intensive environments has increased substantially over time.

Covert channels are used when legitimate communication between the desired endpoint processes is forbidden, e.g. through security mechanisms.  They are also used when the existence of the communication should be kept secret.  This is in contrast to typical methods of securing communication, like cryptography, where the existence of the communication is known, even if the nature of the message is not.  When designing a covert channel, the developer should keep this in mind, and assume that the legitimate channel over which the covert channel operates is under scrutiny.  So, covert channels aim to allow clandestine, secure communication over an otherwise secured and scrutinized resource.

Network covert channels may be classified as storage based or timing based.  Storage based covert channels convey their message by altering part of the legitimate (henceforth {\em carrier}) communications data.  Examples include the aforementioned direct insertion data into an unused or unimportant header field, and the modulation of header fields such as the TCP Initial Sequence Number (ISN) \cite{Goher12}.  Timing based covert channels communicate by modulating some aspect of the carrier communication's timing.  For example, the length of the inter-packet delay (also known as packet inter-arrival time) may be modulated \cite{Cabuk04}, with timing variations conveying the covert message.  Detection of storage based network covert channels relies on the identification of abnormalities in protocol structure or content.  Detection of timing based network covert channels generally relies on statistical analysis of the carrier communication's timing attributes \cite{Cabuk04}, \cite{Cabuk09}.

In this paper we introduce a network covert channel that does not rely on content or protocol-level timing modification of the carrier to send its data, but instead controls the order of {\em named resource} accesses to send its message.  As such, we classify it as a {\em zero-modification storage based network covert channel.}  Our work makes the following contributions to covert channel technology:

{\bf Covert transmission of data through access ordering.}
Whereas traditional network storage based covert channels convey data by changing or adding information to the carrier communication, and typical timing based covert channels operate by modifying the carrier's protocol-level timing, our channel does not modify the carrier content or protocol level timing attributes.  Instead, it modulates the {\em order} of normal, unmodified resource accesses to send its covert message.

{\bf Transaction-level timing modulation for user behavior emulation.}
Timing based covert channels traditionally modify timing attributes that are associated with a deterministic state machine such as a protocol stack.  As such, statistical based detection of such channels may be quite effective.  Our channel, as implemented in this paper, modifies {\em transaction} level timing.  In our case, this is the time between web resource accesses.  Transaction level timing is highly dependent on resource content and on the subjective user.  We can take advantage of this non-deterministic timing structure to effectively emulate user behavior on interactive protocols, avoiding detection through statistical, and in large part behavioral, analysis.

Following this introduction we will proceed to describe the foundational concepts underlying our network covert channel, such as the concept of named resources and the mechanisms of the channel itself.  Included in that description is an overview of our use of transaction timing modulation to emulate user behavior.  Following that, we will outline our web based implementation of the channel, which uses URLs to transmit its covert message.  While our channel is flexible and usable on diverse protocols, we chose an HTTP, or web browsing scenario due to the ubiquity and often innocuous appearance of web associated protocols in a typical network.  These qualities make web protocols excellent {\em camouflage protocols} for our channel.  Following the description of our implementation we will analyze the channel in terms of speed and effectiveness, and finally finish by stating our suggestions for future work, and conclusions.

\section{A Covert Channel Based on Resource Names}

\subsection{Named Resources}
Foundational to our covert channel is the concept of {\em named resources}.  Names may be considered an extension of addresses, generalized to include things that are not necessarily required to access the desired resource, or are not ordinarily considered part of a resource's address proper.  Some examples of names, in the context of this paper, are:

\begin{center}
  {\em http://www.example.org/resource.html}\\
  \vspace{2mm}
  {\em https://example.com/search.php?id=A03F6}\\
  \vspace{2mm}
  {\em smb://fileserver/file.txt}\\
\end{center}

Names exist whether or not they identify or address a resource.  They may be considered separable and could be utilized in piecemeal, if that is helpful.  Names may contain data that identifies a resource, appended with invalid or even nonsensical components.  In our second example above, {\em id} may or not meaningfully associate with the resource {\em search.php}, but exists as part of the name regardless.  Further, {\em search.php} may not actually point to a resource when combined with the given protocol and host.  Names are ubiquitous, existing across protocols and protocol stack layers.  Note that addresses are a subset of names, in our usage of the term.  While we will not make full use of all the attributes of names here, we point them out so that those who might desire to implement a version of our covert channel, perhaps on another protocol, will understand that constraints involving addresses may be ignored when doing so enhances the channel and does not result in enhanced probability of detection by security mechanisms.

Named resource are, of course, resources whose identity (location, etc.) is described by a name.  Our abstraction of names allow our channel to be utilized in a variety of environments on many different layers of a protocol stack.  Though we use URLs as the information bearing names in our implementation here, physical addresses on an 802.11 network, Object Identifiers (OIDs) on a Simple Network Management Protocol (SNMP) server, and many other protocols may conceivably carry information using our covert channel technique.  The important consideration is that the channel should make use of a protocol that allows it to effectively, and most of all securely, communicate in its own particular environment.

It is worthwhile to make mention of the Named Data Networking (NDN) project, a potential candidate for next generation Internet \cite{ndn}.  In the NDN philosophy, everything, down to pieces of content like sections of a video clip, has a name.  So it seems that even next generation networks will be amicable to our covert channel, even more so than today's networks.

\subsection{Formulating a Code-Space from Names}
To convey information using a sequence of names, available names must be associated with codes from a {\em code-space}.  A code-space in this paper is simply a list of codes available to the transmitter.  A simple channel may well make due with a small code-space, consisting, for example, of the lowercase letters a-z, numbers 0-9, and a space character.  Such a channel would be said to have a code-space length of 37.  It may be occasionally necessary to transmit arbitrary binary data such as images, scanned documents, or executables.  To accommodate such a scenario a code-space of length 256 might be used, with each code representing a unique 8-bit byte.  Like the selection of protocol, code-space length is an implementation consideration that is dependent on the particular scenario.

Our channel transmits its covert message by mapping its available names to its desired code-space, and accessing the names in the order required to convey the codes in the message.  It follows that there is a minimum number of names to which the transmitter must have access in order to function.  To maintain the behavioral constraints likely to be required of most secure implementations, the names will need to blend in with normal traffic of the type used in the channel.  For example, in a web based implementation, URLs must be valid, in part so that the transmitter's HTTP query will go off-site (and be seen by the cleverly placed receiver), and in part so that our transactions do not call attention to themselves.  Theoretically, one could utilize the same web resource with different postfixes (e.g. ?id=[random]) to allow for an arbitrarily large code-space.  However, such behavior deviates far from what may be expected from a normal web browsing session, and should be avoided.  The number of names available to the transmitter must be sufficiently high to allow for the code-space to be saturated (i.e. each code has corresponding URLs), and as mapping over a finite code-space may result in uneven mapping, where more URLs map to one code than another, the total number of URLs available to the transmitter should be higher than the number of codes in the code-space.  This is also practical from a behavioral emulation standpoint: repeatedly accessing the same set of resources is unnatural browsing behavior.

It is also possible to transmit more than one code per transaction (e.g. web page access).  As we will see, this significantly improves the data-rate of our channel.  However, increasing the number of codes per transaction significantly increases the number of names required.  The relationship between the minimum number of unique names required of the transmitter, $n$, the code-space length, $l$, and the number of codes sent per transaction, $c$, is given in Equation 1.  This is a minimum number because, as we have stated, we will require many more codes if we must adhere to a non-repetitive, behaviorally constrained series of transactions in our transmitter.  Also, while each name will only associate with one code from the code-space, names do not naturally evenly distribute among the code-space, so frequently more than $n$ names are required to ensure that every code has associated names.

% eqn 1
\begin{equation}
  n = l^c
\end{equation}

We define a {\em code list} as a data structure that stores the list of names associated with a particular code.  To evenly distribute the available names over the code-space, we hash the names before assigning them to a code list.  While the unhashed name is kept in the list, the hash is used in assignment.  The transmitter can make the association between hash (and thus name) and code by, for example, taking the first couple of bytes from the hash, and limiting their maximum value to the code-space length using a modulo operation.  Tables 1 and 2 show an example of such a procedure.  In this case the first two bytes of the hash are taken, allowing two codes of length 256 to be sent per transaction.

\begin{table*}[ht]
 \centering
  \begin{tabular}{| l | l | l |}
    \hline
    {\bf URL} & {\bf Hash} & {\bf Byte} \\
    \hline
    http://www.example.com/news-1.html & 155b1b1ef2eb1781147a97d07ce5c8d298a32d29 & 15 \\
    http://www.example.com/news-2.html & 2bf66e1e1e70fd41e7f2633f71af6819c4783c91 & 2b \\
    http://www.example.com/aboutus.html & d742191535719fece50a8fb7857910a554a4ac0a & d7 \\
    http://www.example.com/news-3.html & 5125b8d9c5f6c338d9ec0bbe952db8a676886f63 & 51 \\
    http://www.example.com/news-4.html & a3825d044dba76a89f6a11e1fae15bdcb6fbf0f7 & a3 \\
    ..... & ..... & .....\\
    \hline
  \end{tabular}
  \begin{center} Table 1: Link to hash pairs\end{center}
\end{table*}

\begin{table*}[ht]
 \centering
  \begin{tabular}{| l | l | }
    \hline
    {\bf Byte to Code} & {\bf URL/character association}\\
    15 mod 25 & http://www.example.com/news-1.html $\leftrightarrow$ {\em v} \\
    2b mod 25 & http://www.example.com/news-2.html $\leftrightarrow$ {\em g} \\
    d7 mod 25 & http://www.example.com/aboutus.html $\leftrightarrow$ {\em 4} \\
    51 mod 25 & http://www.example.com/news-3.html $\leftrightarrow$ {\em h} \\
    a3 mod 25 & http://www.example.com/news-4.html $\leftrightarrow$ {\em p} \\
    ..... & .....\\
    \hline
  \end{tabular}
  \begin{center} Table 2: Hash to character pairs\end{center}
\end{table*}

Sending one of 256 codes per transaction requires a minimum of 256 unique names, minimum because of the possibility of more than one name mapping to one code.  
Sending two codes per transaction requires at least $256^2 = 65,536$ unique names.   Three codes per transaction requires at least $256^3 = 16,777,216$ unique names.
We can see that a smaller code-space implies more efficient channel operation: less names are required to saturate the code lists, or alternatively the same number of names can be used to allow for more than one code to be sent per transaction.  In our web based  implementation, sending two codes per transaction is the practical maximum.  To send three codes per transaction requires that a minimum of $256^3$ URLs be harvested to seed the transmitter.  Such harvesting would be temporally prohibitive.  Even with a lower number of names needed for transmission, obtaining an adequate number may prove challenging.  We have created (and implemented) an effective solution to this problem in our web based channel.  A daemon, running continuously in the background on the user's machine (or another machine, if appropriate) collects URLs as the user browses the web.  Each returned page is examined and links to pages (e.g. links leading to resources ending in .php, .html, etc.) are recorded.  When the transmitter is ready to transmit its message, it takes the list of names from the daemon (which in our case collected upwards of 900 unique links per hour the user was actively browsing), and uses that list in the formulation of the code-space.  Since the daemon is always running, the list of available names continuously grows, and transmitting with bigger code-spaces and more codes per transaction becomes easier as time goes on.

\subsection{Transmitter and Receiver Operation}
Once the transmitter has been {\em primed}, that is, once it has a sufficient number of names available to transmit its covert message, it can begin transmitting.  Once the transmitter is ready, it will transmit as follows.  Every $c$ codes from the message is taken at a time, $c$ being the number of codes sent per transaction.  A name is selected from the code list associated with the code (or codes if $c > 1$).  The selection may be random, as in our simple web based implementation, or adhere to necessary behavioral constraints.  The transmitter simply accesses the name in the typical way (e.g. HTTP GET for a web page), and continues the process until the entire message has been sent.

A universal assumption among covert channels is that the receiver must have eventual access (physical or logical) to the data the transmitter has sent that is relevant to the covert channel.  In our web implementation, the covert channel receiver is normally not the destination of the transmitter's HTTP requests (though such an arrangement would make the receiver's requirement of obtaining the transmitter's web traffic easier).  As such, the transmitter is accessing web resources at diverse locations, for example, from different web servers.  In this case, the receiver should have access to the transmitter's transmission medium at some point before the traffic is separated according to destination.  Given the physical and logical securities in place around many networks (locked communication closets, IPSEC between sites, and so on), realization of this caveat can prove daunting.

While there is no particular method that users of network covert channels employ to give the receiver access to the transmitter's communications, we can provide some potential scenarios as examples.  In the case of corporate or government entities, that have lawful or de facto control over the communication infrastructure of organizations that handle transmitter traffic, intercepting the transmitter's communications is a matter of exercising authority.  This is particularly relevant as covert channels are reserved for the most secret communications, when even the existence of the communication is to be hidden.  So, in this first example, subversive agencies may have illicit access to telecommunications infrastructure, and the transmitter need only innocuously browse the web (in a loose deterministic fashion) in order to securely pass information to them.  Our technique is extensible to other protocols (a few possibilities may include certain remote file access and torrent protocols), so systems with protocol attributes that are inherently public could also provide the receiver with access to the covert transmitter's communications.  Finally, the transmission medium itself may be a broadcast domain, as may be the case in certain radio based systems.  For another, published example of how our system might be integrated, see \cite{Girling87}.

The receiver requires limited knowledge in order to recover the message from the transmitter's transactions.  These are, the hash algorithm used on the names, and the mapping between the hash and the code-space.  The receiver does not need to make assumptions regarding what names are available to the transmitter.  It simply takes note of the names accessed by the transmitter (and their temporal order), hashes them, and correlates the resulting hash to the code-space (similar to what we saw in Tables 1 and 2.)  The decoding can of course be done in real-time or offline (for example, by parsing a tcpdump pcap file with the transmitter's traffic.)

The transmitter and receiver may coordinate the beginning and end of a covert message transmission by a special sequence of codes or by the use of a pre-shared delineation name (e.g. URL).  Using a sequence of codes to delineate the message will require more time than using a delineation name, as multiple codes must be transmitted.  Using a reserved name to mark the end of a transmission may have security implications.  For example, if the transmitter and receiver agree that a certain name marks the end of a transmission, this name may not be used during the transmission of the message without causing the receiver to prematurely determine that the message has ended.  Using this reserved name (or names) at only the beginning and end of the message may have security implications as well, as the content of the name may deviate from what the user is otherwise accessing, and the occurrence of an isolated name among names that are otherwise related (by content or link structure, for example) may be suspicious.  If unrelated traffic exists in the transmitter's message stream (for example, if the transmitting user is browsing the web at the same time the message is being transmitted), similar techniques may be used to differentiate the message traffic from the transmitter's browsing.  If such a technique is used in the web browsing channel implemented here, additional security concerns arise, as the behavioral and timing constraints enforced by the channel for security may be negated by the additional pages the user accesses.

\subsection{Modeling User Behavior}
If the channel's underlying protocol allows for rapid accesses of resources, our channel can be quite fast.  Unfortunately, when user interactive protocols are involved, such as is the case with a web browsing based channel, the time between transactions (read-time) may be high.  In those cases it is not appropriate, from a behavioral emulation perspective, that the transmitter simply access the relevant named resources in sequence without consideration for the natural inter-transaction delay that would occur if a user were actually interacting with those resources.  The covert channel must emulate this delay, to satisfy some degree of behavioral security.

As our channel implementation below uses web resource access to convey its message, we will focus on the timing considerations involved in a web browsing scenario.  In web browsing, the read-time is loosely related to the content of the retrieved resource.  Videos take a while to watch, and games take a while to play.  But, users do not always watch the entire video, or finish the game.  In the case of text base resources, the user may be interested in a long piece of text, and spend a significant amount of time reading it, or they may be bored by it and click away to another page soon after access.  Such dynamics mean that web browsing inter-transaction delays (read-times) are partially, though not entirely, non-deterministic.  The authors of \cite{Liu10} have found that user web read-time can be modeled with some degree of success by examining page attributes such as keyword content, word count, and page geometry, and using this information to formulate a Weibull distribution, from which realistic read-times may be selected.  For tractability, we utilize a simpler approach to determining read-time in our experimental implementation: an exponential distribution (which is a Weibull distribution with a shape parameter of one, and the additional quality of memorylessness), and a formula provided by the Nielsen Corporation that allows us to estimate an average user's read-time when given page word count \cite{Nielsen08}.  The Nielsen equation is shown in Equation 2, where $t_{avg}$ is the average read-time, and $w$ is the number of relevant words.  We consider words that are not part of HTML or code (e.g. JavaScript) as relevant.  We use this average to build an exponential distribution from which to select our pseudo-random read-times.

% eqn 2
\begin{equation}
t_{avg} = 0.44w + 25
\end{equation}

The actual read-time we use in our implementation is taken from a random-exponential distribution that uses $t_{avg}$ as the distribution average.  A uniformly distributed pseudo-random variable $u$ is generated after each transaction and is used in conjunction with the distribution to formulate a read-time, as shown in Equation 3.  We refer to the read-time generation model that employs Equations 2 and 3 as the {\em random-exponential model}.

% eqn 3
\begin{equation}
t_{read} = -t_{avg}\ ln(1 - u)
\end{equation}

By using the random-exponential read-time model in conjunction with the covert channel's accessed resources, we are able to approximate an average web user's behavior from a timing standpoint, allowing our web implementation to blend in, in an ``average'' way.  It is more difficult to emulate a user's {\em resource selection} behavior, however.  In our web implementation the transmitter accesses resources from the relevant code lists pseudo-randomly.  The result is a channel that is able to transmit its message, and emulate a typical user from a timing aspect, but is susceptible to behavioral analysis in terms of resource access ordering.  We have not yet implemented a heuristic to approximate a reasonable resource access ordering, though we have some begun limited investigation into the subject.  The two primary constraining factors determining access ordering in the web browsing case are subject relevance, and link structure relationship.  The typical web browsing user's page accesses are sometimes related by subject, though not always, and subject matter may be distributed across browser tabs and content related to different subject matter accessed in parallel \cite{Weinreich06}.  Further, hyperlink structure is generally coherent; that is, a link in resource {\em A} leads to resource {\em B} which leads to resource {\em C}, regardless of the subject matter.  Subject matter of different sorts may be viewed on a news web page, for example, with link structure leading the user from the news site front page to the various unrelated articles.  A search engine may be used to build hyperlink structure of related subject matter, where a few top level search results are opened in parallel (emulating tabbed browsing), and these pages subsequently traversed.  We have not made sufficient progress into web resource access ordering behavior emulation to include it in our implementation here.  Our channel implementation is sufficient for exposition of its motivating contributions, though it could be improved upon by including access order behavior emulation.

In short, the degree to which the covert channel is able to emulate user behavior in terms of timing, content relevance, and content ordering, determine the level of security of the channel.  Such behavioral constraints are protocol and scenario dependent; we have discussed the considerations that relate to the web browsing case.  Further, keep in mind that imposing restrictions on the channel's access behavior may decrease its data-rate, and increase the time required to prime the system, as the set of available resources for each message code will be constrained not only to those that are relevant by code, but also relevant by resource content.

\subsection{Data-rate}
The data-rate of our covert channel depends on the number of codes sent per transaction, and the time between transactions.  Generally, we can express code-rate as shown in Equation 4, where $r$ is the code-rate in codes per second, $K$ is the number of codes sent per transaction, $d$ is the sum of the network, protocol, and other natural latencies of the system (in seconds), and $t_{read}$ is the read-time inserted between transactions (also in seconds.)  If an implementation is using the random-exponential delay model already described, $t_{read}$ is the same given in Equation 3.  If read-time varies, an average code-rate over a number of transactions $T$ can be expressed as shown in Equation 5.

For implementations using a binary code-space of 256 length, data-rate can also be measured in bits per second, ($bps$).  The $bps$ analogues to Equations 4 and 5 are shown in Equations 6 and 7.

% eqn 4
\begin{equation}
  r = \frac{K}{d+t_{read}}
\end{equation}

%eqn 5
\begin{equation}
  r_{avg} = \frac{(K)(T)}{\sum\limits_{n = 1}^{T} d+t_{read[n]}}
\end{equation}

\begin{equation}
  bps = 8\ \frac{K}{d+t_{read}}
\end{equation}

\begin{equation}
  bps_{avg} = 8\ \frac{(K)(T)}{\sum\limits_{n = 1}^{T} d+t_{read[n]}}
\end{equation}

Our web based channel used two read-time models in testing: the random-exponential model already explained, and a {\em minimum time} model that did not insert any intentional delay between transactions.  In the minimum time model, the delay between transactions is composed entirely of system and network delays.  The tests involving the minimum time delay model yielded much higher data-rates than those using the random-exponential delay model.  This emphasizes a typical trade-off that exists in covert channel design and implementation: speed versus security.  Whereas our minimum time model gives faster data-rates, its unnatural access behavior makes channels using it vulnerable to detection by even naive security systems.

Though we do not explore the topic here, potential data-rate improvements exist for our channel.  These include running multiple channels in parallel (perhaps using different protocols to avoid compromising typical behavioral mechanics of any one protocol), and combining our covert channel with other covert channels.  Of course, the security implications of such {\em hybrid channels} should be well considered before real-world implementation.

\begin{figure*}[ht]
  \centering
    \includegraphics[width=0.85\textwidth]{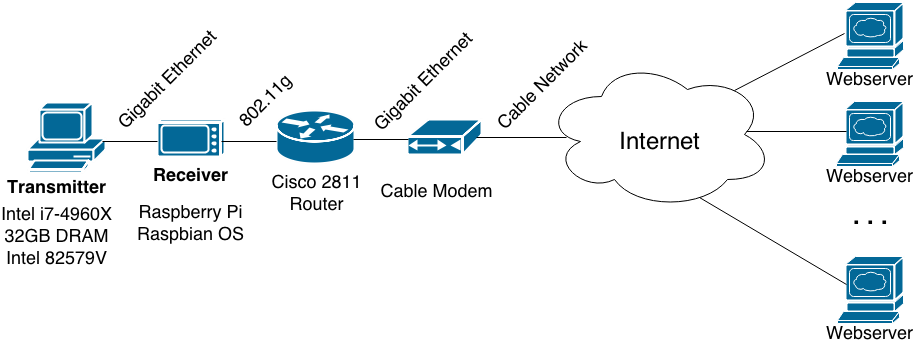}
  \caption{Implementation Scenario}
\end{figure*}

\subsection{Detection}
Security can be applied to communications by employing technologies such as cryptography.  Covert channels are used when evading detection is the primary consideration: when the very existence of the channel should remain a secret.  Techniques used to detect the existence of covert channels have historically focused on behavioral and timing anomalies induced by the channel \cite{Cabuk09} \cite{Goher12}, with timing analysis generally done at the protocol level \cite{Cabuk04} \cite{Gianvecchio11} \cite{Giles02} \cite{Luo08} \cite{Tumoian05}.  As mentioned before, our channel does not change any component of the carrier data.  As such traditional storage based covert channel detection techniques do not apply to our channel.  Also as mentioned, since we do not change protocol level timing of the carrier traffic, our channel is not detectable by typical timing based covert channel detection techniques, which focus on changes in the deterministic timing of protocol stat machines.  We may be deficient from a behavioral standpoint, if an impetus to look for our channel, or those like it, exists.  For example, our inter-transaction read-times as generated by our random-exponential read-time model emulate a ``typical'' user.  It is conceivable, though we believe at the moment unlikely, that variance from the {\em individual} user's read-time behavior may be seen as suspicious.  The under-explored topic of user behavioral emulation in transaction ordering is our most likely vulnerability to detection, though we know of no work that has been done to date on this topic, and simple behavioral emulation measures (such as the search engine based technique mentioned above) is assumed to be adequate for most situations.

\section{Implementation}

Our web-based implementation operates as follows.  A daemon runs continually on the user's machine and observes their web traffic.  For each web resource accessed by the user, the daemon unobtrusively parses the return HTML, adding to its list of available resources (URLs).  For each link that does not already exist in one of its code lists, it assigns it to a list as follows.  First, it hashes the full URL using the SHA-1 hash algorithm.  Then, it takes the first two characters of the hash, modulo the code-space size (which is variable), and assigns the URL to a code list according to the result.  This is similar to the process shown in Tables 1 and 2.  We select the first two bytes of the hashes because the maximum code-space length in our tests is 256, and the maximum number of codes per transaction (i.e. codes per name) we send in the tests is two.  So, by using two bytes, each name can associate with up to one of $256^2$ different codes (see Equation 1.)  In our implementation we ignore behavioral restrictions involving name relation, and consider our channel ready to transmit after all relevant code lists have associated URLs.

The transmitter component of our implementation is tunable in terms of code-space and number of codes sent per transaction.  It can use the random-exponential or minimum time read-time delay models, also configurable by the user. It receives its URL lists from the daemon, who continues adding to its own lists even as the transmitter retrieves pages during transmission.  When directed to transmit the transmitter will send its message $c$ codes at a time, $c$ being the number of codes per transaction as configured for the particular test.  The transmitter continues accessing names until it has finished sending its message.  The transmitter and daemon were programmed in Python and ran on a Linux system (Fedora 20), consisting of commodity hardware including an Intel i7-4960X processor, 32GB of DRAM, and an Intel 82579V Gigabit Ethernet controller.  Both the daemon and transmitter were programmed in Python.

In our implementation the receiver passively forwarded traffic from the transmitter's LAN to the LAN edge router (a Cisco 2811), collecting traffic from the transmitter's IP address in a pcap file with tcpdump.  After transmission, it decoded the transmitter's accesses using roughly the same method the transmitter used to assign its names to code-space (Tables 1 and 2.)  In our experiments the receiver ran on a small, unobtrusive Raspberry Pi, such as might easily go unnoticed hidden in a communications closet or above a dropped ceiling.  The Pi ran Raspbian Linux, and the receiver software was coded in Python.  For each test, both the transmitter and receiver scripts were configured for the particular code-space mapping.  Aside from the hash algorithm and transmitter IP address (so that it could sort the transmitter from the aggregate communication stream), no further information was required of the receiver to decode the message.

The topology of the test network is shown in Figure 1.  Note that the destination servers exist in diverse locations on the Internet, and so their associated hardware and software are variable.

\vspace{1cm}
\section{Experiments}
Several tests were executed that allowed us to observe the trade-offs between different system parameters in our system.  The parameters varied were message length (16, 32, and 64 bytes), code-space size (32, 128, and 255 codes), codes per transaction (1 or 2 codes), and delay model (random-exponential or minimum delay.)  These parameters were chosen as they illustrate the adaptability of the channel to different scenarios (e.g. code-space lengths), and to better understand how they each relate to channel performance.  Forty tests were run for configurations using the random-exponential read-time model, 20 were run for those using the minimum delay model.

\section{Results}
The results of the individual test configurations are shown in Table 3, where the read-time model, given in the {\em delay} column, is either MT for minimum time, or RE for random-exponential. Unsurprisingly (and as mentioned before), the read-time model has the most significant impact on the channel performance.  For those tests using the random-exponential delay model, message length had minimal impact on the channel speed, code-space length had minimal impact as well, though increasing the number of codes sent per transaction from one to two resulted in a substantial (about four-fold) increase in data-rate.  The extent to which the delay model had an effect on channel data-rate is shown in Figure 2a.  Figures 2b through 2d convey visually how message length, code-space size, and codes per transaction affected channel rate, with both delay models included in the calculations.  Figures 3a through 3c show this same information taken from only the random exponential read-time delay model tests.

\begin{table*}[ht]
  \centering
  \begin{tabular}{| c | c | c | c | c | c |}
    \hline
    C.S. Len. & Msg. Len. & Codes / Hash & Delay & Avg. (bps) & Std. Dev.\\ \hline \hline
    255 & 16 bytes & 1 & MT & 3.61 & 14.10 \\ \hline
    255 & 16 bytes & 1 & RE & 0.90 & 0.30 \\ \hline
    255 & 16 bytes & 2 & MT & 2.94 & 20.41 \\ \hline
    255 & 16 bytes & 2 & RE & 0.46 & 0.81 \\ \hline
    255 & 32 bytes & 1 & MT & 2.92 & 10.07 \\ \hline
    255 & 32 bytes & 1 & RE & 0.09 & 0.40 \\ \hline
    255 & 32 bytes & 2 & MT & 11.23 & 14.75 \\ \hline
    255 & 32 bytes & 2 & RE & 0.39 & 0.67 \\ \hline
    255 & 64 bytes & 1 & MT & 3.12 & 15.27 \\ \hline
    255 & 64 bytes & 1 & RE  & 0.08 & 0.64 \\ \hline
    255 & 64 bytes & 2 & MT  & 14.57 & 69.00 \\ \hline
    255 & 64 bytes & 2 & RE  & 0.35 & 1.04 \\ \hline
    128 & 16 bytes & 1 & MT & 3.23 & 15.94 \\ \hline
    128 & 16 bytes & 1 & RE & 0.09 & 0.28 \\ \hline
    128 & 16 bytes & 2 & MT & 11.78 & 26.40 \\ \hline
    128 & 16 bytes & 2 & RE & 0.45 & 0.71 \\ \hline
    128 & 32 bytes & 1 & MT & 3.29 & 15.67 \\ \hline
    128 & 32 bytes & 1 & RE  & 0.87 & 0.35 \\ \hline
    128 & 32 bytes & 2 & MT & 12.67 & 40.38 \\ \hline
    128 & 32 bytes & 2 & RE & 0.42 & 0.78 \\ \hline
    128 & 64 bytes & 1 & MT & 2.80 & 6.47 \\ \hline
    128 & 64 bytes & 1 & RE  & 0.08 & 0.45 \\ \hline
    128 & 64 bytes & 2 & MT  & 16.47 & 70.62 \\ \hline
    128 & 64 bytes & 2 & RE & 0.41 & 1.69 \\ \hline
    32 & 16 codes  & 1 & MT & 2.42 & 13.46 \\ \hline
    32 & 16 codes  & 1 & RE & 0.88 & 0.33 \\ \hline
    32 & 16 codes  & 2 & MT & 9.01 & 22.94 \\ \hline
    32 & 16 codes  & 2 & RE & 0.52 & 0.74 \\ \hline
    32 & 32 codes  & 1 & MT & 2.49 & 18.05 \\ \hline
    32 & 32 codes  & 1 & RE & 0.09 & 0.42 \\ \hline
    32 & 32 codes  & 2 & MT & 10.39 & 31.03 \\ \hline
    32 & 32 codes  & 2 & RE & 0.38 & 0.84 \\ \hline
    32 & 64 codes  & 1 & MT & 3.04 & 18.78 \\ \hline
    32 & 64 codes  & 1 & RE & 0.08 & 0.39 \\ \hline
    32 & 64 codes  & 2 & MT & 10.28 & 64.4 \\ \hline
    32 & 64 codes  & 2 & RE & 0.37 & 1.12 \\ \hline
  \end{tabular}
  \begin{center}Table 3: Test Results \end{center}
\end{table*}

\section{Future Work and Conclusions}

We have taken a novel approach to creating a network covert channel that by its nature evades typical covert channel detection techniques such as those that employ content analysis to detect storage based covert channels, and those that use statistical analysis to detect timing based channels.   Our channel is flexible and may be implemented using arbitrary protocols, as best suits the particular situation.

The most challenging aspect of this covert channel is user behavior emulation for transaction selection.  Not only should resources be accessed in a natural way, but the number of resources requires to make such natural accesses available may be quite large.  In the web case, filling the code lists with enough URLs to avoid repetition may take a significant amount of time (which is why we employ a daemon to prime the system before and between transmissions); adding the caveat that resources must be related (by content, or hyperlink structure) increases the required number of resources dramatically.

Research shown in \cite{Liu10}, \cite{Huang10}, and \cite{Weinreich06} demonstrate that the Nielsen formula for computing average read-time based on page word count is over simplified.  To better emulate user timing behavior, our channel would need to include the results developed in those sources.

It is also worth looking into the viability of using other protocols in our channel, as other protocols may be more amicable to our channel concepts in certain environments.  For example, remote file access protocols such as Server Message Block (SMB) or Network File System (NFS) protocols may be natural carriers for our channel in certain environments.

\begin{figure*}[ht!]
    \centering
        \subfigure[Avg. Rate by Delay Model]{
            \label{fig:first}
            \includegraphics[width=0.3\textwidth]{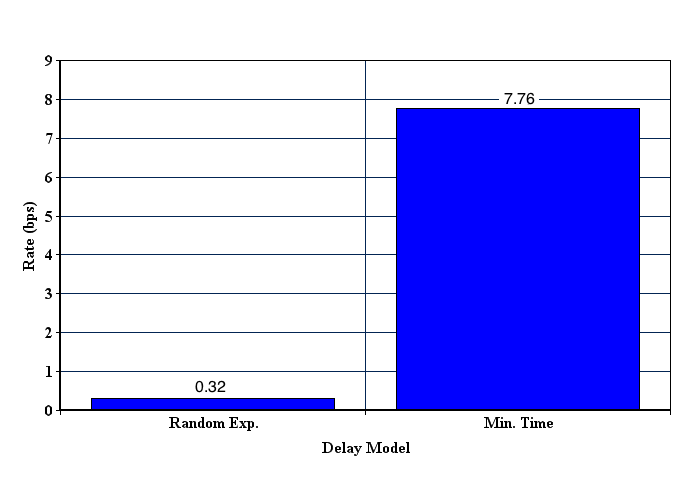}
        }
        \subfigure[Avg. Rate by Message Length]{
           \label{fig:second}
           \includegraphics[width=0.3\textwidth]{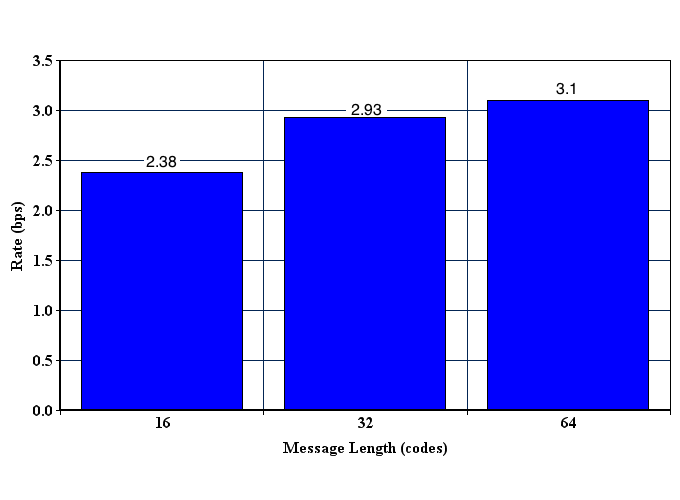}
        }\\ %  ------- End of the first row ----------------------%
        \subfigure[Avg. Rate by Code-Space Length]{%
            \label{fig:third}
            \includegraphics[width=0.3\textwidth]{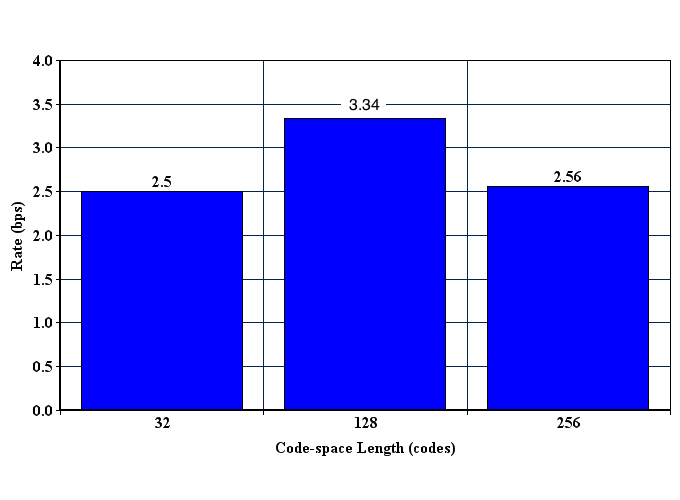}
        }
        \subfigure[Avg. Rate by Codes per Transaction]{
            \label{fig:fourth}
            \includegraphics[width=0.3\textwidth]{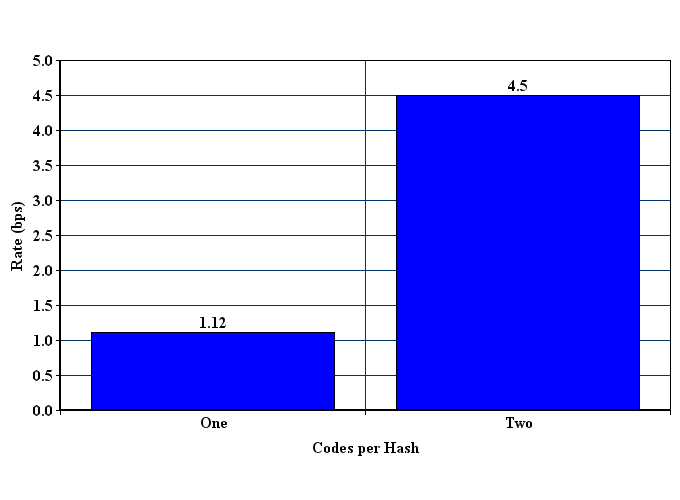}
        }
    \caption{
        Average rates among all tests
     }
   \label{fig:subfigures}
\end{figure*}

\begin{figure*}[ht!]
    \centering
        \subfigure[Avg. Rate by Message Length (Random Exp.)]{
            \label{fig:first}
            \includegraphics[width=0.3\textwidth]{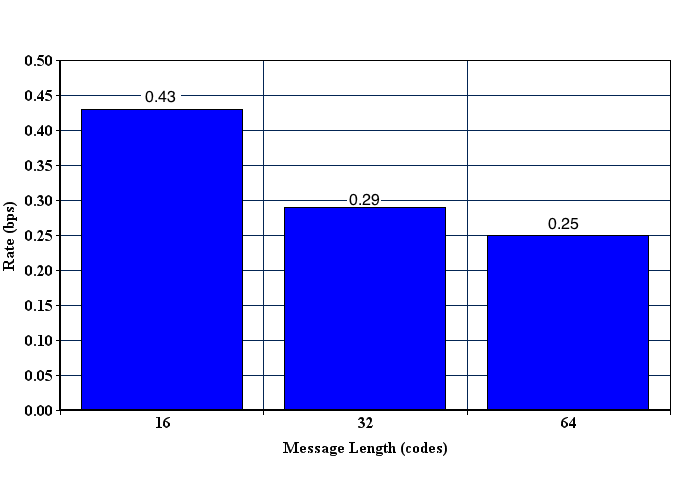}
        }
        \subfigure[Avg. Rate by Code-Space Length (Random Exp.)]{
           \label{fig:second}
           \includegraphics[width=0.3\textwidth]{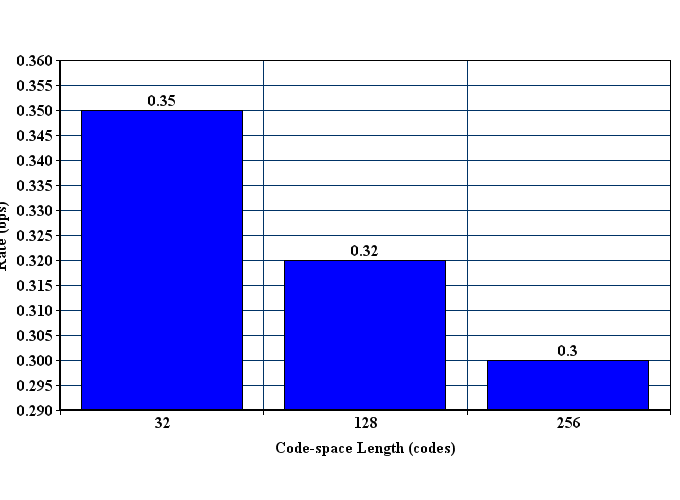}
        }\\ %  ------- End of the first row ----------------------%
        \subfigure[Avg. Rate by Codes per Hash (Random Exp.)]{%
            \label{fig:third}
            \includegraphics[width=0.3\textwidth]{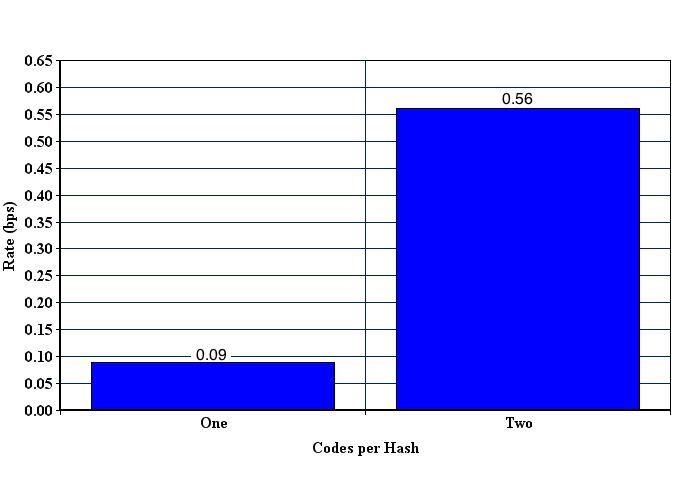}
        }
    \caption{
        Average rates among random exponential delay model tests
     }
   \label{fig:subfigures}
\end{figure*}

\clearpage

\bibliographystyle{abbrv}
\bibliography{channel}

\begin{thebibliography}{10}

\bibitem{ndn}
Named data network.
\newblock \url{http://named-data.net/}.
\newblock Accessed 2014-05-14.

\bibitem{Ahmadzadeh13}
S.~Ahmadzadeh and G.~Agnew.
\newblock Turbo covert channel: An iterative framework for covert communication
  over data networks.
\newblock In {\em INFOCOM, 2013 Proceedings IEEE}, pages 2031--2039, 2013.

\bibitem{Cabuk04}
S.~Cabuk, C.~E. Brodley, and C.~Shields.
\newblock Ip covert timing channels: Design and detection.
\newblock In {\em Proceedings of the 11th ACM Conference on Computer and
  Communications Security}, CCS '04, pages 178--187, New York, NY, USA, 2004.
  ACM.

\bibitem{Cabuk09}
S.~Cabuk, C.~E. Brodley, and C.~Shields.
\newblock Ip covert channel detection.
\newblock {\em ACM Trans. Inf. Syst. Secur.}, 12(4):22:1--22:29, Apr. 2009.

\bibitem{Gasior11}
W.~Gasior and L.~Yang.
\newblock Network covert channels on the android platform.
\newblock In {\em Proceedings of the Seventh Annual Workshop on Cyber Security
  and Information Intelligence Research}, CSIIRW '11, pages 61:1--61:1, New
  York, NY, USA, 2011. ACM.

\bibitem{Gianvecchio11}
S.~Gianvecchio and H.~Wang.
\newblock An entropy-based approach to detecting covert timing channels.
\newblock {\em Dependable and Secure Computing, IEEE Transactions on},
  8(6):785--797, 2011.

\bibitem{Gilbert09}
P.~Gilbert and P.~Bhattacharya.
\newblock An approach towards anomaly based detection and profiling covert
  tcp/ip channels.
\newblock In {\em Information, Communications and Signal Processing, 2009.
  ICICS 2009. 7th International Conference on}, pages 1--5, 2009.

\bibitem{Giles02}
J.~Giles and B.~Hajek.
\newblock An information-theoretic and game-theoretic study of timing channels.
\newblock {\em Information Theory, IEEE Transactions on}, 48(9):2455--2477,
  2002.

\bibitem{Girling87}
C.~Girling.
\newblock Covert channels in lan's.
\newblock {\em Software Engineering, IEEE Transactions on}, SE-13(2):292--296,
  1987.

\bibitem{Goher12}
S.~Goher, B.~Javed, and N.~Saqib.
\newblock Covert channel detection: A survey based analysis.
\newblock In {\em High Capacity Optical Networks and Enabling Technologies
  (HONET), 2012 9th International Conference on}, pages 057--065, 2012.

\bibitem{Huang10}
J.~Huang and R.~W. White.
\newblock Parallel browsing behavior on the web.
\newblock In {\em Proceedings of the 21st ACM Conference on Hypertext and
  Hypermedia}, HT '10, pages 13--18, New York, NY, USA, 2010. ACM.

\bibitem{Liu10}
C.~Liu, R.~W. White, and S.~Dumais.
\newblock Understanding web browsing behaviors through weibull analysis of
  dwell time.
\newblock In {\em Proceedings of the 33rd International ACM SIGIR Conference on
  Research and Development in Information Retrieval}, SIGIR '10, pages
  379--386, New York, NY, USA, 2010. ACM.

\bibitem{Luo08}
X.~Luo, E.~Chan, and R.~Chang.
\newblock Tcp covert timing channels: Design and detection.
\newblock In {\em Dependable Systems and Networks With FTCS and DCC, 2008. DSN
  2008. IEEE International Conference on}, pages 420--429, 2008.

\bibitem{Luo12}
X.~Luo, E.~Chan, P.~Zhou, and R.~Chang.
\newblock Robust network covert communications based on tcp and enumerative
  combinatorics.
\newblock {\em Dependable and Secure Computing, IEEE Transactions on},
  9(6):890--902, 2012.

\bibitem{Mazurczyk13}
W.~Mazurczyk, K.~Szczypiorski, and J.~Lubacz.
\newblock Four ways to smuggle messages through internet services.
\newblock {\em Spectrum, IEEE}, 50(11):42--45, 2013.

\bibitem{Nielsen08}
J.~Nielsen.
\newblock How little do users read?, 2009.

\bibitem{Tumoian05}
E.~Tumoian and M.~Anikeev.
\newblock Network based detection of passive covert channels in tcp/ip.
\newblock In {\em Local Computer Networks, 2005. 30th Anniversary. The IEEE
  Conference on}, pages 802--809, 2005.

\bibitem{Weinreich06}
H.~Weinreich, H.~Obendorf, E.~Herder, and M.~Mayer.
\newblock Off the beaten tracks: exploring three aspects of web navigation.
\newblock In {\em Proceedings of the 15th international conference on World
  Wide Web}, pages 133--142. ACM, 2006.

\end{thebibliography}

\end{document}